\documentclass[pra,aps,10pt,amsmath,amssymb,twocolumn,superscriptaddress]{revtex4-2}

\usepackage{graphicx}
\usepackage{dcolumn}
\usepackage[dvipsnames,table,xcdraw]{xcolor}
\usepackage{bm}
\usepackage[colorlinks=true]{hyperref} 
\usepackage[export]{adjustbox}
\usepackage{braket}

\usepackage{ragged2e}
\usepackage{subcaption}

\newcommand{\edfig}[4]{
\begin{figure*}
    \centering
    \refstepcounter{figure}
    \includegraphics[width=#2]{#1}
    \captionsetup{justification=Justified}
    \caption*{\textbf{Extended Data Fig.~\thefigure:} #3}
    \label{#4}
\end{figure*}
}

\begin{document}

\title{Doppler-induced tunable and shape-preserving frequency conversion\\of microwave wave packets}

\author{Felix Ahrens}
\thanks{These authors contributed equally to this work}
\email{fahrens@fbk.eu}
\affiliation{Fondazione Bruno Kessler (FBK), 38123, Trento, Italy}

\author{Enrico Bogoni}
\thanks{These authors contributed equally to this work.}
\affiliation{Fondazione Bruno Kessler (FBK), 38123, Trento, Italy}
\affiliation{Department of Physics, University of Milano-Bicocca, 20126, Milano, Italy}

\author{Renato Mezzena}
\affiliation{Department of Physics, University of Trento, 38123, Trento, Italy}
\affiliation{TIFPA - INFN, Trento, 38123, Italy}

\author{Andrea Vinante}
\affiliation{CNR-IFN, 38123, Trento, Italy}
\affiliation{Fondazione Bruno Kessler (FBK), 38123, Trento, Italy}

\author{Nicol\`o Crescini}
\affiliation{Fondazione Bruno Kessler (FBK), 38123, Trento, Italy}

\author{Alessandro Irace}
\affiliation{Fondazione Bruno Kessler (FBK), 38123, Trento, Italy}
\affiliation{Department of Physics, University of Milano-Bicocca, 20126, Milano, Italy}

\author{Andrea Giachero}
\affiliation{Department of Physics, University of Milano-Bicocca, 20126, Milano, Italy}
\affiliation{INFN - Milano Bicocca, 20126 Milano, Italy}
\affiliation{Bicocca Quantum Technologies (BiQuTe) Centre, 20126 Milano, Italy}

\author{Gianluca Rastelli}
\affiliation{Pitaevskii BEC Center, CNR-INO and Department of Physics, University of Trento, 38123 Trento, Italy}

\author{Iacopo Carusotto}
\affiliation{Pitaevskii BEC Center, CNR-INO and Department of Physics, University of Trento, 38123 Trento, Italy}

\author{Federica Mantegazzini}
\affiliation{Fondazione Bruno Kessler (FBK), 38123, Trento, Italy}
\affiliation{Department of Physics, University of Trento, 38123, Trento, Italy}
\affiliation{TIFPA - INFN, Trento, 38123, Italy}

\begin{abstract}
In superconducting electronics, the ability to control the frequency of microwave wave packets is crucial for several applications, such as the operation of superconducting quantum processors and the readout of superconducting sensors. We introduce a new approach to microwave frequency conversion that harnesses a dynamic Doppler effect induced by a propagating front that separates regions of different phase velocities. Employing a high-kinetic-inductance superconducting transmission line in a travelling-wave geometry, we were able to implement frequency shifts of microwave wave packets at 500\,MHz and 4\,GHz of up to 3.7\,\% while fully preserving their temporal shape. In contrast to conventional methods based on frequency-mixing, our Doppler-induced frequency-conversion method avoids spurious mixing products, is continuously tunable by a quasi-dc current amplitude, and allows to imprint arbitrary patterns on the instantaneous frequency profile of temporally long wave packets. By engineering transmission lines that allow for larger phase-velocity changes and/or by cascading multiple Doppler-induced frequency conversions, an unlimited amount of frequency shifting is in principle attainable. These features demonstrate the potential of our frequency-conversion technique as a promising tool for advanced control of microwave wave packets for different quantum applications. 
\end{abstract}
                           
\maketitle 

%
The Doppler effect is a fundamental wave phenomenon in which the observed frequency varies according to the relative velocity between the source of the wave and the detector \cite{Doppler1842, Ballot1845}.
An intriguing variant of this effect arises when two linear media with different wave phase velocities ($v_1, v_2$) are separated by an interface moving at a constant and high speed $v$~\cite{Gafar2019}. 
In this configuration, the moving boundary acts as a dynamic receiver that scatters waves at the same frequency in its rest frame. This leads to marked Doppler shifts in the frequencies of the scattered waves as seen from the laboratory frame.
In particular, while the frequency of the transmitted wave remains unchanged in the rest frame of the moving interface as in standard refraction, in the laboratory frame an incident wave with frequency $\omega_1$ propagating in the medium with phase velocity $v_1$ undergoes a Doppler shift upon traversing the moving interface. The resulting transmitted wave in the second medium, characterized by the phase velocity $v_2$, is shifted to the frequency $\omega_2 = \omega_1+\Delta\omega$, which can be straightforwardly calculated~\cite{Gaburro:06,gaburro2008photonic}, predicting
\begin{equation}
\label{eq:frequencyshift}
\frac{\omega_2}{\omega_1}
= \frac{1-v/v_1}{1-v/v_2}
\end{equation}
and, most remarkably, anticipating that the shape of arbitrary wave packets is neither altered nor deformed. This suggests that this Doppler-induced frequency-conversion process would allow for a continuous and precise tuning of the frequency of a wave packet without distorting its profile, or significantly attenuating its intensity, or spoiling its quantum coherence.

Although it is challenging to move physical objects at sufficient speed to observe a sizable frequency-shift, a moving interface between two media of different phase velocities can be virtually induced by spatiotemporally modulating the refractive index of a single medium~\cite{Morgenthaler:1958} with a second propagating pulse via some optical nonlinear effect~\cite{Gafar2019,Gaburro:06,gaburro2008photonic}. This strategy is conceptually related to dynamic photonic structures~\cite{Yanik2005-DynamPhotStruct} and time-refraction~\cite{mendoncca2024time}, and has been exploited to observe frequency conversion and pulse-compression effects in lightwave photonics~\cite{Geltner2002,Upham_2010,Kampfrath:PRA2010,Kondo:PRL2014,Castellanos:PRA2015,Kondo.PhysRevA.97,Apffel:PRL2022,schiff2021front}.
As all of these experiments are based on a spectral analysis of the converted wave packet, they do not specifically address the issue of pulse shape preservation.

In this work, we use a superconducting medium to demonstrate a tunable Doppler-induced frequency-conversion effect in the microwave domain.
Microwave photonics is gaining importance because it combines the low-loss, long-range advantages of microwave radiation with the extreme spatial confinement and design flexibility of microfabricated waveguides and structures \cite{Nori2017}.
A primary factor driving the rapid expansion of this field is that superconducting qubits \cite{Devoret2013,blais2021circuit}, spin qubits \cite{Zhang2018, Burkard.RevModPhys.95}, and hybrid solid-state quantum devices \cite{Xiang2013, Kurizki2015} inherently operate in the microwave regime, making microwave photonics a natural and indispensable platform for their precise control and manipulation.

Our experiments are based on a superconducting high-kinetic-inductance microwave transmission line whose inductance can be dynamically tuned by a current running through the line's centre conductor. In this way, the phase velocity of the transmission line can be spatiotemporally controlled through a travelling current front. Thus, it is possible to shift the frequency of a counter-propagating wave packet that meets the current front inside the transmission line.
With our proof of concept experiment, we demonstrate Doppler-induced frequency shifting of arbitrarily shaped wave packets in the microwave domain, and we explicitly show continuous tunability of the shift via the current intensity. Going beyond the intrinsic limitations of cavity-based frequency-conversion schemes~\cite{Sandberg2008b,Xu:PRApp2022}, we specifically demonstrate that our frequency-conversion process preserves the shape of the wave packet and is robust against variations of the shape of the current front. We finally show how our scheme also allows for the design of complex instantaneous frequency patterns within a single, temporally long wave packet.

\begin{figure*}[ht]
    \centering
    \begin{subfigure}{\textwidth}
        \centering
        \includegraphics[width=0.8\linewidth]{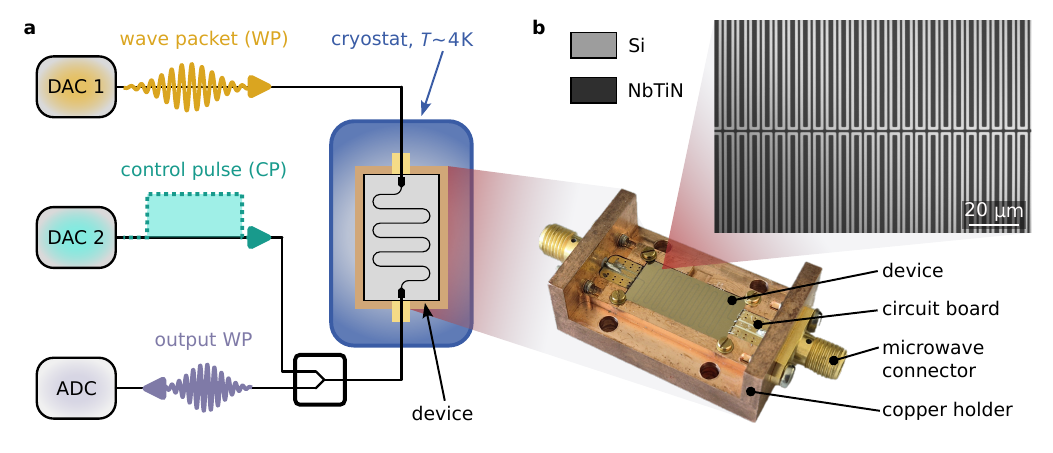}
        \label{fig:sub1}
    \end{subfigure}
    \vfill
    \begin{subfigure}{\textwidth}
        \centering
        \includegraphics[width=0.8\linewidth]{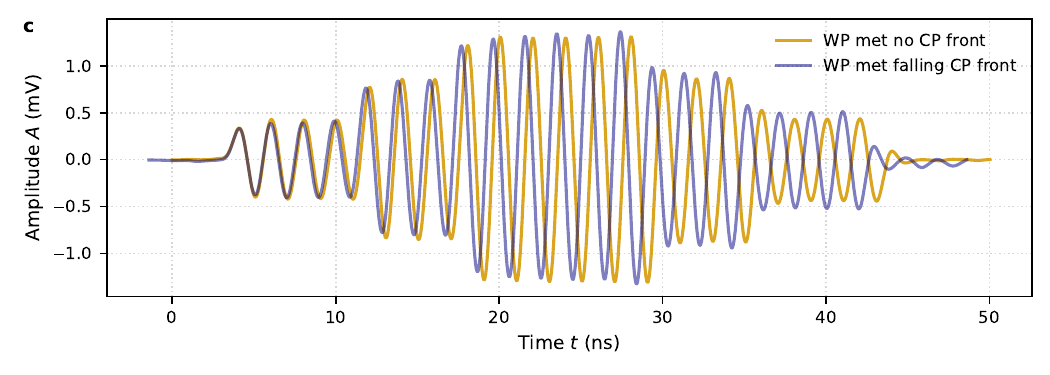}
        \label{fig:sub2}
    \end{subfigure}
    \captionsetup{justification=Justified}
    \caption{\textbf{Experimental setup and concept.} \textbf{a}, schematic outline of the experimental setup. Two DACs (digital-to-analogue converters) are employed to create a microwave wave packet (WP) and a control pulse (CP), which are routed to the opposite ports of a superconducting high-kinetic-inductance transmission line and counter-propagate through it. The outgoing wave packet is routed through a splitter to an ADC (analogue-to-digital converter). \textbf{b}, cryogenic microwave setup: the superconducting device is packaged in a copper holder equipped with printed circuit boards and microwave connectors. A scanning electron microscopy (SEM) image shows the microscopic structure of the superconducting transmission line, implemented as a coplanar waveguide loaded with finger-shaped capacitor elements, patterned in a thin NbTiN film on a Si substrate \cite{DW2023}. \textbf{c}, measured output wave packets in absence (gold line) and presence (blue line) of a control pulse for identical input wave packets. The carrier frequency of the first wave packet coincides with that of the input wave packet, i.e.~500\,MHz, while the carrier frequency of the second wave packet is visibly increased due to the Doppler-induced frequency-conversion effect induced by the falling front of the control pulse.}
    \label{FIG:Fig1_setup}
\end{figure*}

\section*{Concept and experimental implementation}

At the core of our experimental setup lies a superconducting microwave waveguide consisting of a coplanar artificial transmission line, a scanning electron micrograph of which is presented in Fig.\,\ref{FIG:Fig1_setup}\textbf{c}. The device is patterned into a high-kinetic-inductance film based on the superconductor NbTiN \cite{DW2023} and has a length equivalent to $\sim40\,$ns of propagation time \cite{Ahrens2024}.
In such a device, the kinetic inductance per unit length $L_\mathrm{k}$ can be controlled through the bias current $I$, varying at first order according to the relation $L_\mathrm{k}(I) = L_0 \cdot (1 + I^2/I_*^2 + \mathcal{O}(I^4))$, where $L_0$ is the kinetic inductance per unit length at zero current and $I_*$ is the nonlinearity scale \cite{Parmenter1962, Anthore2003, Eom2012}. In turn, the phase velocity $v_\mathrm{p}$ will depend on the bias current:

\begin{equation}
    v_\mathrm{p} \approx \frac{1}{\sqrt{L_0 C}} \left( 1-\frac{I^2}{2I_*^2} + \mathcal{O}(I^4) \right)\,,
    \label{eq:propagationvelocity}
\end{equation}

\noindent where $C$ is the capacitance per unit of length of the transmission line and geometric contributions to the line inductance have been neglected. 

To spatiotemporally vary the phase velocity $v_\mathrm{p}$ of the device, we inject a control pulse consisting of a rising current front, followed by a plateau and a falling front. The two moving fronts dynamically divide the transmission line into regions of different phase velocities. Based on the general arguments sketched above, a microwave wave packet that encounters a moving current front inside the device is then expected to experience a frequency shift according to Eq.~\eqref{eq:frequencyshift}, while maintaining the shape of its temporal envelope. 

Our experimental setup to demonstrate this Doppler-induced frequency-shifting effect is sketched in Fig.~\ref{FIG:Fig1_setup}\textbf{a}.
The microwave wave packet and the control pulse are generated by two digital-to-analogue converters (DACs) and are routed to opposite ports of the high-kinetic inductance transmission line, which is housed inside a cryostat with a base temperature of $\sim4\,$K. 
The microwave wave packet exiting the device is routed to an analogue-to-digital converter (ADC), where it is recorded and analyzed.

As a first proof-of-principle experiment, using an arbitrary-waveform generator as DAC and an oscilloscope as ADC, we generate a wave packet with a staircase-like envelope and a carrier frequency of 500\,MHz, and we record the full time-trace of the output wave packet, as shown in Fig.\,\ref{FIG:Fig1_setup}\textbf{c}.
The gold line indicates a waveform recorded when no control pulse has been applied, while the blue line indicates the waveform of a wave packet that has encountered the falling front of a control pulse: it exhibits a frequency shift of 14\,MHz, while its temporal profile is preserved.
A detailed description of the experimental setup is provided in the Methods section and in Extended Data Fig.\,\ref{FIG:experimental_setup_extended}\textbf{a}.

\section*{Frequency shift measured by frequency scan}

Going beyond this proof-of-principle experiment, we demonstrate and study the Doppler-induced frequency-shifting effect in the gigahertz-frequency regime, which is particularly relevant to the control of superconducting qubits. Both the generation of wave packet and control pulse, and the recording of the output wave packet are from now on handled by an FPGA-based RF system on chip (RFSoC), where the wave packet and the control pulse are synthesized directly by the DACs of the RFSoC, while the output wave packet is digitized and digitally down-converted at frequency $\omega_\mathrm{d}$ \textit{in situ} by the ADC of the RFSoC. Further details on this experimental apparatus are available in the Methods section and in Extended Data Fig.\,\ref{FIG:experimental_setup_extended}\textbf{b}.

\begin{figure*}[ht]
\centering
\includegraphics[width=0.8\textwidth, trim= 48 25 45 60, clip]{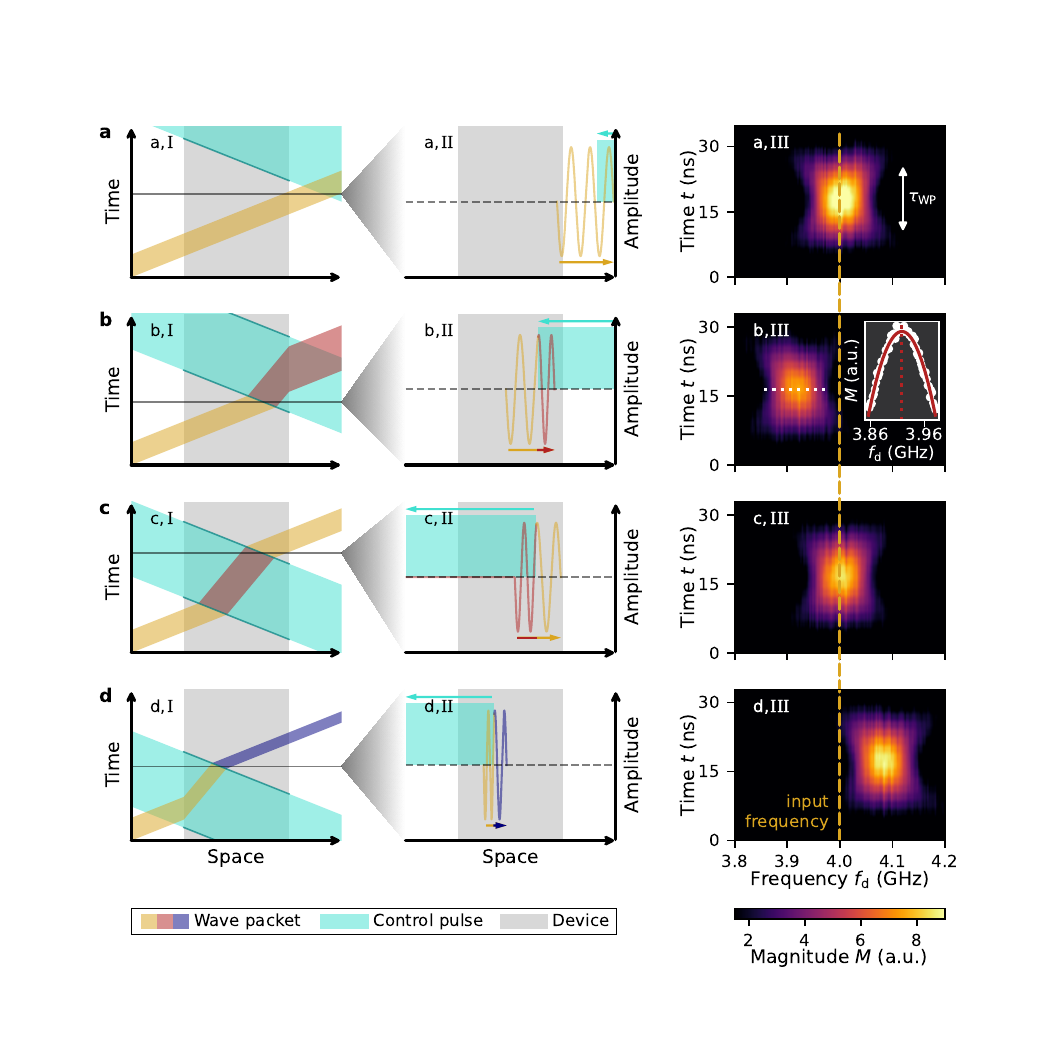}
\captionsetup{justification=Justified}
\caption{\textbf{Frequency shifting dynamics.} The dynamics of the frequency-conversion experiment is shown for four configurations \textbf{a}, \textbf{b}, \textbf{c} and \textbf{d}, corresponding to different delays between the wave packet (WP) and the control pulse (CP): \textbf{a}, WP and CP do not meet inside the device. 
\textbf{b}, 
WP meets the rising front of CP inside the device and undergoes a redshift. 
\textbf{c}, WP meets both rising front and falling front of CP inside the device. The respective red- and blueshift of the WP cancel out. 
\textbf{d}, WP meets the falling front of CP inside the device and undergoes a blueshift.
Column \textbf{I}: spacetime diagrams showing the evolution of the WP and the CP for the four configurations. The red and blue colour of the originally gold WP indicate the spacetime regions in which it has been red- or blue-shifted, respectively.
Column \textbf{II}:
snap-shots of the WP and CP relative positions in space at a given instant in time. 
The arrows indicate the respective direction of movement.
Column \textbf{III}: experimentally measured magnitude of the wave packet at the output of the device plotted as colourmap for the four configurations sketched in \textbf{I} and \textbf{II}, demonstrating the expected frequency shift with respect to the input frequency.
$\tau_\mathrm{wp}$ is the pulse durations of the WP. The white curve in the inset of \textbf{b,III} shows the data along a fixed-time cut indicated as a dotted white line in the main plot. The red solid line is a quadratic fit to the data, while the red dashed line marks the centre of the parabola, i.e.~the output frequency $\omega_\mathrm{out}/(2\pi)$.
Details on the measurement are discussed in the main text.
}\label{FIG:Fig2_freq_shift_amp}
\end{figure*}

We study the frequency-conversion effect under four different conditions (Fig.\,\ref{FIG:Fig2_freq_shift_amp}\textbf{a}\,--\,\textbf{d}), which we outline in the spacetime diagrams (Fig.\,\ref{FIG:Fig2_freq_shift_amp}, column (\textbf{I})), and in the fixed-time sections thereof (Fig.\,\ref{FIG:Fig2_freq_shift_amp}, column (\textbf{II})). Each condition corresponds to a different delay between the wave packet and the control pulse. 
Condition (\textbf{a}) shows the case in which the wave packet travels entirely through the device before the control pulse reaches the device. Consequently, the frequency of the wave packed is not altered. 
In condition (\textbf{b}), the wave packet encounters the rising front of the control pulse inside the device and exits the device before meeting the falling front of the control pulse. Due to the Doppler-induced frequency-shifting effect, the outgoing wave packet is redshifted. Condition (\textbf{c}) captures the situation in which the wave packet interacts with both the rising and the falling front of the control pulse. Here, the wave packet is temporarily redshifted inside the device as a result of the encounter with the rising front. However, the interaction with the falling front causes a blueshift that exactly cancels out the previous redshift, so that the outgoing wave packet exhibits the same frequency as the ingoing one. In condition (\textbf{d}), the wave packet enters the device after the rising front of the control pulse has exited the device. The wave packet meets the falling front inside the device, which blueshifts its frequency. 

To experimentally implement these four different situations, we inject a $\tau_\mathrm{WP}=15$\,ns long wave packet with a carrier frequency of $\omega_\mathrm{in}/(2\pi)=4.0$\,GHz and a rectangular envelope into one port of the device. Simultaneously, into the other port of the device, we inject a counter-propagating rectangular control pulse with a pulse duration of $\tau_\mathrm{CP}=30$\,ns and a peak amplitude of $1.62\,\mathrm{mA}$. By changing the delay between their respective arrival to the device, we make the wave packet interact with none (\textbf{a}), either (\textbf{b},\textbf{d}), or both fronts (\textbf{c}) of the control pulse.

We measure the outgoing wave packet for each of the four conditions by recording its time trace 200 times, digitally down-converting it each time at a different frequency $f_\mathrm{d}=\omega_\mathrm{d}/(2\pi)$ between 3.8\,GHz and 4.2\,GHz, and digitally low-pass filtering it with a cut-off frequency $f_\mathrm{LP}=42\,\mathrm{MHz}$. From the baseband quadratures $I(f_\mathrm{d}, t)$ and $Q(f_\mathrm{d}, t)$, we determine the magnitude $M(f_\mathrm{d}, t)=\sqrt{I(f_\mathrm{d}, t)^2+Q(f_\mathrm{d}, t)^2}$, which is shown as colourmaps in Fig.\,\ref{FIG:Fig2_freq_shift_amp}, column (\textbf{III}). The bright areas indicate the times and down-conversion frequencies for which the output wave packet is detected. Naturally, because of the finite pulse-length $\tau_\mathrm{WP}$, the spectral content of the wave packet is spread over a frequency range $\sim1/\tau_\mathrm{WP}$ around the carrier frequency. Furthermore, the rectangular temporal envelope of the wave packet gives rise to an additional frequency broadening at the beginning and end of the pulse, causing the butterfly shape in the colourmaps.

As expected for the Doppler-induced frequency-shifting effect, under conditions (\textbf{b}) and (\textbf{d}), we observe that the frequency band of the wave packet is, respectively, redshifted and blueshifted from the input frequency, while in cases (\textbf{a}) and (\textbf{c}) the frequency remains unchanged. To quantify the amount of frequency shifting, we fit the midmost part of the central fixed-time cut of the wave packet with a parabola \footnote{The measured frequency profile of the wave packet depends on both the spectral components corresponding to the finite pulse length of the wave packet and on the characteristics of the low-pass filter used during the digital down-conversion. As the profile is expected to be symmetric, for the sake of simplicity, we extract the position of the maximum by a quadratic fit.}, as indicated in the inset of Fig.\,\ref{FIG:Fig2_freq_shift_amp}\textbf{b},\textbf{III}. We assign the frequency corresponding to the maximum of the parabola to be the output frequency $\omega_\mathrm{out}$ of the wave packet, and define the global frequency shift of the wave packet to be $\Delta\omega_\mathrm{glob}=\omega_\mathrm{out}-\omega_\mathrm{in}$. In this way, we find the redshift under condition (\textbf{b}) to be $\Delta\omega_\mathrm{glob,b}=-81.8(3)$\,MHz and the blueshift under condition (\textbf{d}) to be $\Delta\omega_\mathrm{glob,d}=81.1(3)$\,MHz.

\section*{Frequency shift measured through phase evolution}

\begin{figure}[ht!]
\centering
\includegraphics[width=0.98\linewidth]{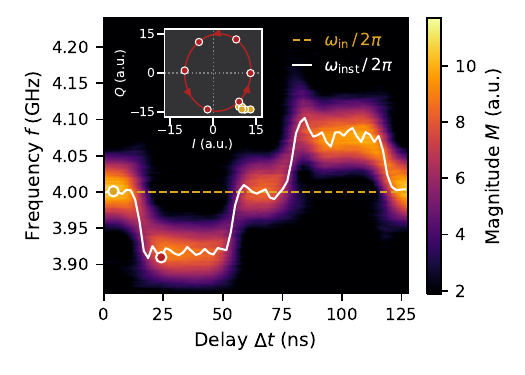} \\
\captionsetup{justification=Justified}
\caption{\textbf{Instantaneous frequency shift.} Merged data obtained from 70 individual $\tau_\mathrm{WP}=15\,$ns long wave packets with an initial carrier frequency $\omega_\mathrm{in}/(2\pi)=4\,\mathrm{GHz}$ which have encountered a $\tau_\mathrm{CP}=40\,$ns long control pulse with an amplitude $1.58\,\mathrm{mA}$ for different delays between $0-128\,$ns. 
    The output wave packets have been digitally down-converted at 191 different frequencies between 3.86\,GHz and 4.24\,GHz and the resulting magnitude is plotted as colourmap.
    For each delay an instantaneous frequency shift at a single point inside the wave packet has been determined from the phase evolution. The combined data from all wave packets has been superimposed on the colourmap as a white line.
    The inset shows the phase evolution in the $I$--$Q$ plane for two output wave packet digitally down-converted at $\omega_\mathrm{in}$, one of which underwent no frequency shift (yellow markers) and one of which was redshifted (red markers). As a guide to the eye we have overlaid red arrows in the latter case. 
    }
\label{FIG:Fig3}
\end{figure}

An alternative way to measure the frequency shift of the wave packet can be achieved by restricting oneself to digitally down-convert the output wave packet with a single frequency only, which we choose to be the input frequency of the wave packet $\omega_\mathrm{d}=\omega_\mathrm{in}$. The baseband signal can then be represented by its two quadratures $I$ and $Q$ as a function of time.
If the output wave packet's frequency $\omega_\mathrm{out}$ coincides with the down-conversion frequency $\omega_\mathrm{in}$, the baseband signal appears stationary in the $I$--$Q$ plane, as illustrated by the yellow data in the inset of Fig.\,\ref{FIG:Fig3}. 
If, instead, the output wave packet's frequency $\omega_\mathrm{out}(t)$ differs from the down-conversion frequency $\omega_\mathrm{in}$, a beating can be observed, which manifests itself in a rotation of the baseband output signal in the $I$--$Q$ plane, described by the phase angle 
\begin{equation}
     \varphi(t) = 
     \int^{t}_{t_0}\!\!\! \mathrm{d}t'
\left[
\omega_\mathrm{in} - \omega_\mathrm{out}(t')
\right]  \,,
\end{equation}
which is illustrated by the red data in the inset of Fig.\,\ref{FIG:Fig3}. We infer the time-resolved frequency shift directly from the phase evolution of the baseband signal as
\begin{equation}
    \Delta\omega_\mathrm{inst}(t) = - \frac{\mathrm{d} \varphi(t) }{\mathrm{d}t}\,.
    \label{EQ:freq_shift_from_dphidt}
\end{equation}

In Fig.\,\ref{FIG:Fig3} we verify the agreement between our two different methods for evaluating the frequency shift, namely the measurements at varying down-conversion frequency explained before and the measurements at fixed down-conversion frequency and derivative of the phase explained here. 
For this, we reconstruct all possible encounter conditions between a wave packet and a control pulse by merging the data of 70 individual measurements, each with a unique delay in the range $\Delta t=0-128\,$ns.
The wave packet was always prepared with an initial carrier frequency $\omega_\mathrm{in}/(2\pi)=4.0\,\mathrm{GHz}$, a rectangular envelope, and a pulse length of $\tau_\mathrm{WP}=15$\,ns, while the control pulse exhibited a rectangular shape, an amplitude of 1.58\,$\mathrm{mA}$ and a pulse length of $\tau_\mathrm{CP}=40\,\mathrm{ns}$. 

To construct the colourmap for our down-conversion-frequency-scan method, for each delay $\Delta t$ we have measured the magnitude map $M_{\Delta t}(f_\mathrm{d}, t)$ of the corresponding output wave packet by sweeping the down-conversion frequency in steps of 2\,MHz analogously to the data shown in Fig.\,\ref{FIG:Fig2_freq_shift_amp}. We then determine the central fixed-time profile $M_{\Delta t}(f_\mathrm{d}, t=t_\mathrm{centre})$ of the output wave packet for a time $t_\mathrm{central}$ corresponding to the centre of the output wave packet. Finally, we combine the fixed-time profiles for all delays $\Delta t$ into the final magnitude map $M(\Delta t, f_\mathrm{d})$.

For the phase-evolution method, we digitally down-convert the output wave packets only at $\omega_\mathrm{d}=\omega_\mathrm{in}$. For each delay, we determine the instantaneous frequency shift $\Delta\omega_\mathrm{inst}(\Delta t)$ through Eq.~\eqref{EQ:freq_shift_from_dphidt} well inside the wave packet as detailed in the Methods section. We superimpose $\Delta\omega_\mathrm{inst}(\Delta t)$ for all delays as a white line on the colourmap in Fig.\,\ref{FIG:Fig3}.

\begin{figure*}[ht!]
\centering
\includegraphics[width=0.8\linewidth]{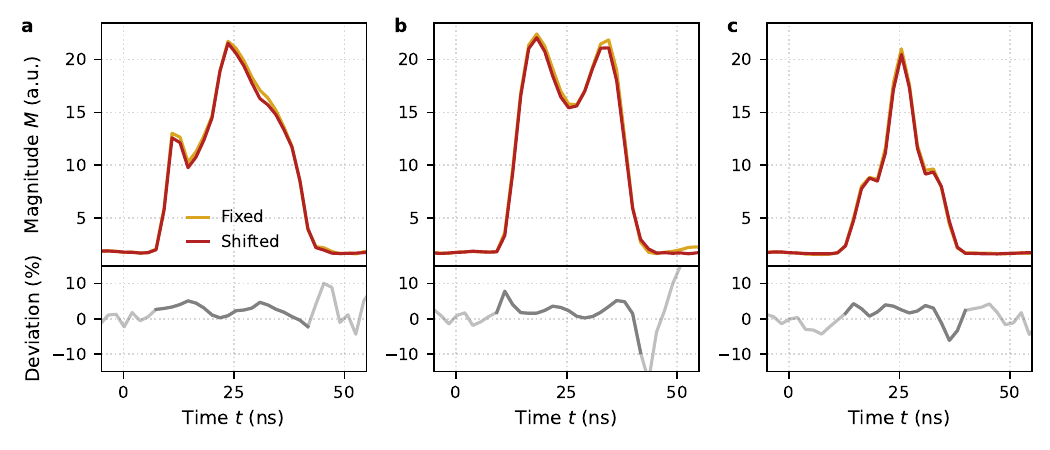}
\captionsetup{justification=Justified}
\caption{\textbf{Pulse shape preservation.} Measured envelopes of wave packets with initial frequency $\omega_\mathrm{in}/(2\pi)=4$\,GHz, pulse length $\tau_\mathrm{WP}=30$\,ns and non-trivial temporal envelopes. The wave packets indicated in gold (red) have passed through the device in absence (presence) of a rising front of height $0.52\,\mathrm{mA}$. Digital down-conversion was performed at the expected output frequencies, respectively. In the lower panels the relative difference between the fixed-frequency and the shifted-frequency pulses is plotted.}
\label{FIG:Fig4_shape_preservation}
\end{figure*}

With both measurement methods, a redshift is observed when the wave packet interacts with the rising front of the control pulse only, i.e.~for delays $16\,\mathrm{ns}<\Delta t<56\,\mathrm{ns}$; a blueshift is found when the wave packet interacts with the falling front of the control pulse only, i.e.~for delays $78\,\mathrm{ns}<\Delta t<118\,\mathrm{ns}$; for delays $56\,\mathrm{ns}<\Delta t<78\,\mathrm{ns}$ the wave packet encounters both fronts and the experienced red- and blueshifts cancel out exactly. It should be noted that the transition between regions characterized by a frequency shift and those without is not infinitely sharp; rather, it exhibits a finite gradient, as evidenced, for instance, near $\Delta t\approx16\,\mathrm{ns}$. This finite slope in the frequency shift is caused by the finite slope of the control-pulse front, as sketched in Extended Data Fig.~\ref{FIG:partial_lifting}. We investigate this effect in more detail and discuss its consequences below.

\section*{Pulse-shape preservation}
Correspondingly to the frequency shift of the wave packet arising from the Doppler-induced frequency-shifting effect, the temporal envelope of the output wave packet also experiences an expansion or contraction with respect to the input wave packet. Aside from this contraction or expansion effect, it is naturally expected that the pulse shape is preserved upon crossing the counter-propagating current front. 

In Fig.\,\ref{FIG:Fig4_shape_preservation}, we evaluate the effect of the frequency shift on the pulse shape for three 30\,ns long wave packets with initial carrier frequencies of $\omega_\mathrm{in}/(2\pi)=4.0\,\mathrm{GHz}$ and different non-trivial envelopes. We have measured the temporal amplitude profile of the output wave packets by digitally down-converting them with their carrier frequency $\omega_\mathrm{d}=\omega_\mathrm{out}$, which was determined from their baseband phase evolutions in a separate measurement before. 

In the upper panels, the gold lines represent the temporal amplitude profiles of wave packets that moved through the device in the absence of any control pulse. As they do not experience any frequency-shift effect, their envelopes act as reference. 
Wave packets traversing a rising control-pulse front of height $0.52\,\mathrm{mA}$ inside the device result in redshifted output wave packets whose measured envelopes are visualized as red lines in Fig.\,\ref{FIG:Fig4_shape_preservation} and show good agreement with the reference gold lines. 

Given the relatively small frequency shift of 0.25\,\%, the expansion of the wave packet due to the Doppler shift can in fact be neglected. The pulse-shape deviation between the output wave packets at initial and redshifted carrier frequencies can be quantified as the relative difference between the curves that describe the envelopes, which is plotted in the lower panels of Fig.\,\ref{FIG:Fig4_shape_preservation} indicating deviations $\leq10\,\%$. These minor deviations arise from two effects: on the one hand, a comparably low sampling rate of $0.55\,$Gs/s after decimation makes the comparison more susceptible to jitter and timing inaccuracies, increasing the relative deviation, particularly at steep slopes. On the other hand, small impedance mismatches inside the device caused by the presence of the control pulse can cause internal reflections. While the former effect is measurement-related and can easily be solved by measuring with a faster ADC, the latter one can be mitigated through a dedicated engineering of the transmission-line geometry.

\section*{Amplitude-controlled frequency shift}
One of the main advantages of the frequency-conversion mechanism that we present here is its precise control by amplitude tuning.
To demonstrate this feature, we have measured the phase evolution of wave packets with an initial frequency of $\omega_\mathrm{in}/(2\pi)=4.0\,\mathrm{GHz}$ that have interacted with the rising front of control pulses of different amplitudes $I_\mathrm{CP}=0.08-2.03\,\mathrm{mA}$. For each amplitude $I_\mathrm{CP}$, we have digitally down-converted the output wave packets at frequency $\omega_\mathrm{d}=\omega_\mathrm{in}$ and we have recorded the unwrapped phase $\varphi$ of $24$ output wave packets with different creation delays $-3.6\,\mathrm{ns}<\Delta t<38.2\,\mathrm{ns}$ between the wave packet and the control pulse. 

\begin{figure}[ht]
\centering
\includegraphics[width=0.98\linewidth]{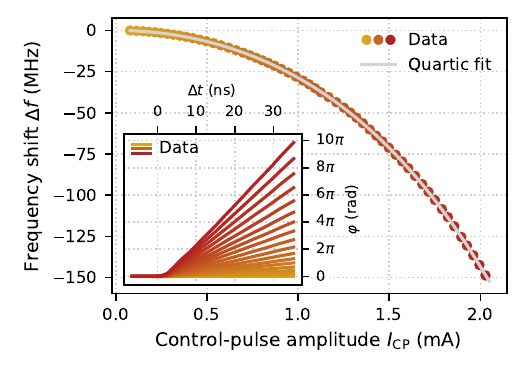}
\captionsetup{justification=Justified}
\caption{\textbf{Amplitude-controlled frequency shift.} The inset shows the unwrapped measured phase $\varphi$ of wave packets with an initial frequency of $\omega_\mathrm{in}/(2\pi)=4.0\,\mathrm{GHz}$ that have interacted with the rising front of a control pulse and that have consecutively been digitally down-converted at $\omega_\mathrm{d}=\omega_\mathrm{in}$. Data are plotted for control-pulse amplitudes between $I_\mathrm{CP}=0.08\,\mathrm{mA}$ (lightest orange line) and $I_\mathrm{CP}=2.03\,\mathrm{mA}$ (darkest red line). Each line represents the phase data for a fixed $I_\mathrm{CP}$ and has been constructed by merging the data obtained from 24 wave packets with different delays $\Delta t$ with respect to the control pulse. The main plot shows the frequency shift obtained from the slope of the phase data in the inset. The frequency shift was fitted with Eq.\,\eqref{eq:shift_from_current} taking into account both quadratic and quartic contributions in $I_\mathrm{CP}$. The fit is shown as light grey line.}
\label{FIG:Fig5}
\end{figure}

The inset of Fig.\,\ref{FIG:Fig5} illustrates the combined phase data for all delays $\Delta t$ and control-pulse amplitudes $I_\mathrm{CP}$. For each value of the amplitude $I_\mathrm{CP}$, the phase starts to deviate from zero for delays $\Delta t>0\,\mathrm{ns}$, which corresponds to the wave packet encountering the rising front of the control pulse. Well after this onset of the phase evolution, the slope of the phase evolution turns linear. The negative of the slope be identified with the wave packets' frequency shift $\Delta\omega$. We plot this frequency shift as a function of the control-pulse amplitude as shown in the main panel of Fig.\,\ref{FIG:Fig5}. For the maximum control-pulse amplitude of $2.03\,\mathrm{mA}$, which is still well below the critical current of $I_\mathrm{c}=2.5\,\mathrm{mA}$ of the superconducting NbTiN structures of the device, we report a frequency shift of $-149$\,MHz, corresponding to a carrier frequency shift of $-3.7\,\%$. Additionally, in Extended Data Fig.\,\ref{FIG:shift_oscilloscope}, we present data for wave packets with a carrier frequency of 500\,MHz, which have been shifted by up to $4.1\,\%$.

Based on Eqs.\,\eqref{eq:frequencyshift} and \eqref{eq:propagationvelocity}, we expect that the redshift $\Delta\omega(I_\mathrm{CP})$ caused by a rising control-pulse front of height $I_\mathrm{CP}$ can be expressed in the form

\begin{equation}
    \Delta\omega(I_\mathrm{CP})=-\frac{\omega_\mathrm{in}}{4}\left(\frac{I_\mathrm{CP}}{I_*}\right)^2+\mathcal{O}\left(I_\mathrm{CP}^4\right)\,,
    \label{eq:shift_from_current}
\end{equation}

\noindent where $|v_\mathrm{CP}|=|v_\mathrm{WP}(I_\mathrm{CP}=0)|$ has been assumed. The solid line in the main panel of Fig.\,\ref{FIG:Fig5} is a fit to the experimental data based on Eq.\,\eqref{eq:shift_from_current} also including a quartic contribution in $I_\mathrm{CP}$. The extracted value $I_*=6.15\,\mathrm{mA}$ of the scaling current is compatible with independent results that we obtained from bias-current-dependent transmission measurements of the device, indicating a good agreement between our frequency-shift model and the data.

\section*{Instantaneous frequency modulation}

While so far we have discussed the global frequency shift of the wave packet as a whole upon the interaction with a single or two infinitely sharp control-pulse fronts, a more subtle physics is found when generalising the mechanism to a single long wave packet that interacts with a series of fronts of different amplitudes. 

\begin{figure*}[ht]
    \centering
    \includegraphics[width=0.8\textwidth]{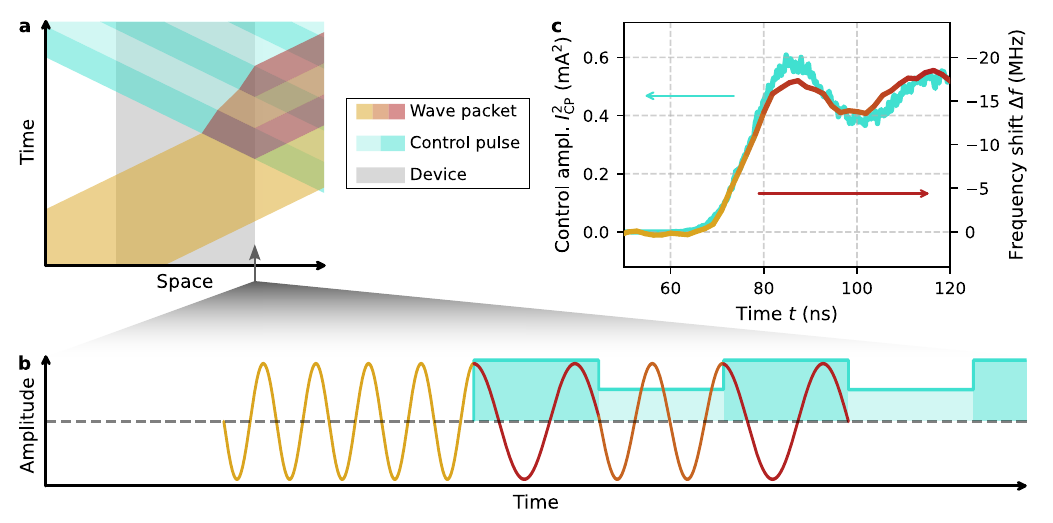}
    \captionsetup{justification=Justified}
    \caption{\textbf{Instantaneous frequency modulation.}
    \textbf{a}, theoretical spacetime diagram of an input wave packet (WP) 
    encountering a series of control-pulse (CP) fronts of different amplitudes inside the device. Each rising CP front leads to a redshift of the WP frequency, while each falling front leads to a blueshift. As a result, the output WP contains segments of varying frequency as highlighted in
    \textbf{b}, where the theoretical temporal profiles of both the CP and the modulated WP amplitudes are sketched. \textbf{c}, measured time-resolved frequency shift for a WP with an initial frequency of 4\,GHz subjected to a CP with the non-trivial temporal profile indicated in turquoise. 
    }
    \label{FIG:Fig6}
\end{figure*}

In Fig.\,\ref{FIG:Fig6}\textbf{a} we sketch a spacetime diagram of a wave packet that interacts with a series of rising and falling current fronts of varying amplitudes. In Fig.\,\ref{FIG:Fig6}\textbf{b}, we highlight a fixed-position cut that visualizes the time evolution of both the control pulse and the wave packet at the output of the device. The first segment of the wave packet has travelled through the device without having encountered any control-pulse front, and therefore, its frequency is unchanged. The second segment of the wave packet has interacted with a rising front and is consequentially redshifted. The third segment of the wave packet has interacted not only with the rising front but also with a minor falling front, which partially compensates for the redshift. Therefore, the frequency of the third segment is lower than the input frequency, but not as low as the frequency of the second segment. The fourth and last segment of the wave packet features again the same redshift as the second segment since it has interacted with yet another rising front.

Moving from the case of multiple discrete fronts with different amplitudes to the limit of smooth control wave forms, it should be possible to shift each time point $t$ of a wave packet to an individual instantaneous frequency $\omega_\mathrm{inst}(t)$. If a point in the wave packet has encountered the front edge of the control pulse inside the device, its instantaneous output frequency $\omega_\mathrm{inst}$ will be determined by the value $I_\mathrm{CP}$ of the amplitude that the control pulse takes at the device output port at the moment when that specific point in the wave packet leaves the device.

In Fig.\,\ref{FIG:Fig6}\textbf{c}, we present the measurement of the instantaneous frequency shift $2\pi\Delta f_\mathrm{inst}(t)=\omega_\mathrm{inst}(t)-\omega_\mathrm{in}$ of a wave packet that has encountered a smooth control waveform. We have digitized the applied control waveform and depicted it as turquoise data points.
The input wave packet has been created with a carrier frequency $\omega_\mathrm{in}/(2\pi)=4\,\mathrm{GHz}$ and a rectangular envelope. We have digitally down-converted the output wave packet at frequency $\omega_\mathrm{d}=\omega_\mathrm{in}$ to deduce the instantaneous frequency shift from the phase evolution of the baseband signal
via Eq.\,\eqref{EQ:freq_shift_from_dphidt}.

The striking agreement between the instantaneous frequency shift of the wave packet and the shape of the applied control waveform proves that, in fact, the shape or steepness of the control-pulse front do not affect the frequency shift process. It is rather the value of the instantaneous control-pulse amplitude at the device output port that dictates the instantaneous frequency shift of each point in the wave packet leaving the device.

This finding has advantageous practical consequences for two use cases: Firstly, the instantaneous nature of the frequency shift allows the creation of arbitrary instantaneous frequency patterns $\omega_\mathrm{out}(t)$ within the microwave wave packet by applying a control-pulse amplitude modulation $I_\mathrm{CP}(t)$ determined by Eq.\,\ref{eq:shift_from_current}. For instance, a down-chirp $\omega_\mathrm{out}(t)=\omega_\mathrm{in}-\frac{1}{2}kt^2$ can in first order be implemented with a linearly increasing control pulse $I_\mathrm{CP}=I_*\sqrt{2k/\omega_\mathrm{in}}\,t$, as sketched in Extended Data Fig.\,\ref{FIG:partial_lifting}. 

Secondly, in the case of a global frequency shift of the wave packet, as, for instance, required in quantum-processor control, the frequency shift is determined only by the plateau value of the control-pulse amplitude after the front. Therefore, our frequency-conversion method is completely insensitive to variations and nonidealities in the front shape caused, for example, by limitations on the front-rise time of the electronic apparatus.

\section*{Conclusions and outlook}

We have experimentally investigated a frequency-conversion method based on a front-induced dynamic Doppler effect, and we have demonstrated it for the first time in a superconducting transmission line in the microwave domain. 
For this, we exploited a superconducting device based on a high-kinetic-inductance transmission line with a travelling-wave geometry, where the local phase velocity is dynamically tuned through a travelling current front: the interaction between a propagating microwave wave packet and a counter-propagating rising (falling) current front is shown to result in a redshift (blueshift) of the microwave wave packet.

This Doppler-induced frequency-conversion method offers high control and tunability of the frequency shift of a wave packet over a wide frequency range while ensuring a precise preservation of the pulse shape. We have directly verified each of these features with dedicated experiments, showing 
(i) the preservation of non-trivial shapes of wave packets in the gigahertz regime, 
(ii) the tunability of the frequency shift via the control-pulse amplitude, 
(iii) the insensitivity of the frequency shift to the current front shape in case of global frequency changes, and
(iv) the possibility to imprint arbitrary patterns on the instantaneous frequency profile of a temporally long wave packet via the instantaneous amplitude of the control pulse.

From a quantitative point of view, we have measured a frequency shift of up to 3.7\,\% at 4\,GHz, which translates into a few hundreds of megahertz for carrier frequencies in the band 4\,--\,8\,GHz. The theoretically achievable maximum frequency shift for high-kinetic inductance structures is in fact limited to 5\,\% by the direct link between the scaling current $I_*$ and the critical current $I_\mathrm{c}$~\cite{Zhao2022}. This limitation can be straightforwardly overcome by using other examples of transmission-line geometries with dynamically tunable local phase velocities, such as Josephson-Junctions-based metamaterials. The dependence of the phase velocity of a transmission line based on a chain of Josephson Junctions with critical current $I_\mathrm{c}$ on the bias current $I$ is given by $v_\mathrm{ph}\propto[1-(I/I_\mathrm{c})^2]^\frac{1}{4}$, which theoretically allows for frequency shifts up to 100\,\%. For large phase velocity jumps, specific attention will have to be paid to reflections due to impedance mismatches and nonlinear distortions. While these issues can be solved with a specific design of the transmission-line properties, an alternative way to achieve arbitrarily large frequency shifts is to simply cascade multiple devices operated with dedicated control-pulse sequences.

In conclusion, we would like to highlight how the physics of our Doppler-induced frequency-conversion mechanism is fundamentally different from conventional frequency mixing. By relying on a Doppler-shift-induced frequency-conversion effect at a phase-velocity front rather than a standard nonlinear frequency-mixing effect, spurious mixing products are inherently avoided, and the frequency shift is amplitude-controlled instead of frequency-controlled. This means that no external coherent source tuned to the frequency difference is needed, and the frequency shift is continuously tunable over wide ranges. 
Although these features are common to any implementation of our method in any frequency domain, other useful features are characteristic of our microwave implementation. For instance, the cryogenic line carrying the microwave signal can be heavily filtered, approaching a very narrow frequency bandwidth, and thus significantly suppressing noise. Being based on a superconducting microwave circuit, this approach also offers full compatibility with cryogenic environments, and calibration protocols do not need to be executed during operation. Our proof-of-concept experiment demonstrates how the presented frequency-conversion approach can represent an attractive alternative to conventional frequency mixing methods, with advantages that are particularly relevant for applications such as the control and readout of quantum-processors \cite{Leonard2019, Demarets2025, Pernas2026}. Taking inspiration from on-going work in the lightwave context~\cite{matsuda2016deterministic, Fenwick2026}, a natural next step will be to experimentally confirm that our frequency-conversion method fully preserves the quantum-coherence properties of microwave wave packets.

\section*{Methods}
\textbf{Device:} The device is based on a superconducting microwave waveguide consisting of a stub-loaded coplanar high-kinetic-inductance artificial transmission line patterned into an 11\,nm-thick NbTiN film with a kinetic inductance of $\sim10\,$pH/sq \cite{DW2023, Ahrens2024}. The film was deposited by reactive magnetron-sputtering and the transmission line was patterned by SF$_6$-based plasma etching. The phase velocity in the transmission line is less than 2\,\% of the vacuum speed of light and the total travel time from the input port to the output port of the device is $\sim40\,$ns.

\textbf{Experimental setups:} To measure the Doppler-induced frequency conversion of microwave wave packets, we have employed two different setups: A first setup that enables the measurement of full time-traces of wave packets with carrier frequencies up to 800\,MHz, and a second setup that allows for the measurement of baseband traces of wave packets with carrier frequencies up to 8\,GHz.

\textbf{Experimental setup for frequency-conversion experiments at 500\,MHz:} 
We create both the wave packet and the control pulse with the two channels of an arbitrary-wave generator, which ensures high temporal control of the delay $\Delta t$ between the two pulses. The carrier frequency of the wave packet is chosen to be $\omega_\mathrm{in}/(2\pi)=500\,$MHz, its envelope is arbitrary and non-trivial, and its length is set to $\tau_\mathrm{WP}=40\,$ns, while the envelope of the control pulse is rectangular and its length is $\tau_\mathrm{CP}=100\,$ns. Through coaxial cables, the wave packet and the control pulse are routed to opposite ports of the device which is cooled to a temperature of $T=4.2\,$K by immersion in a liquid helium bath. The outgoing wave packet is routed through a wideband power splitter to the input of an oscilloscope where its full time-trace is recorded. The oscilloscope is synchronized with and triggered by the arbitrary-wave generator. The experiment is repeated with a rate of 250\,kHz to acquire sufficient statistics. The experimental setup is sketched in Extended Data Fig.\,\ref{FIG:experimental_setup_extended}\textbf{a}.

\textbf{Experimental setup for frequency-conversion experiments at 4\,GHz:} 
Both the wave packet and the control pulse are created by the two DAC channels of an FPGA-based RFSoC system, ensuring high temporal control over the creation delay $\Delta t$ \cite{Qibosoq2026}.
The wave-packet exhibits an arbitrary and non-trivial shape and is created with a carrier frequency of $\omega_\mathrm{in}/(2\pi)=4\,$GHz, while the control pulse is created with a rectangular or arbitrary shape.
The wave packet and the control pulse are routed to the opposite ports of the device, which is housed inside a cryocooler with base-temperature $T\approx4\,$K.
To divide the signal path of the control pulse and the output wave packet, we employ a diplexer. The spectral components of the control pulse fall into the pass band of the low-pass port of the diplexer, i.e.~DC\,--\,3\,GHz, while the spectral components of the output wave packet fall into the pass band of the high-pass port, i.e.~4\,--\,20\,GHz, which allows for an effective and distortion-free separation of the signal paths.
We finally route the outgoing wave packet to the ADC channel of the FPGA, where it is down-converted and recorded with an effective sample rate of $~0.55\,$GS/s after decimation. A sketch of the experimental setup is shown in Extended Data Fig.\,\ref{FIG:experimental_setup_extended}\textbf{b}

\textbf{Procedure for measuring $\omega_\mathrm{inst}(\Delta t)$:} In principle, the measurement of the instantaneous frequency $\omega_\mathrm{inst}$ as a function of the delay $\Delta t$ shown in Fig.\,\ref{FIG:Fig3} could have been carried out by making a single $\tau_\mathrm{WP}\geq 125\,\mathrm{ns}$ long wave packet interact with a $\tau_\mathrm{CP}=40\,\mathrm{ns}$ long control pulse inside the device. In this case, $\Delta t$ denotes the delay between the control pulse and a single point inside the wave packet. The instantaneous frequency change at $\Delta t$ can be determined according to Eq.~\eqref{EQ:freq_shift_from_dphidt}. To reduce the impact of statistical fluctuations, the measurement can be repeated $N$ times and, consequently, the instantaneous frequency can be calculated as
$\omega_\mathrm{inst}(\Delta t)=\sum_{i=1}^N\omega_{\mathrm{inst,}i}(\Delta t)/N$
where $\omega_{\mathrm{inst,}i}$ denotes the measured instantaneous frequency at the $i-$th repetition. However, the length of the wave packet $\tau_\mathrm{WP}=125\,\mathrm{ns}$ would be significantly longer than the propagation time through the device of $\tau_\mathrm{p}\sim40\,\mathrm{ns}$, leading to reflections inside the device and consequently interference. To mitigate the impact of internal reflections, we utilize wave packets of a length of $\tau_\mathrm{WP}=15\,\mathrm{ns}<\tau_\mathrm{p}$. To cover the full range of possible encounter conditions between the wave packet and the control pulse, we repeat the measurement 70 times, each time with a unique temporal distance $0\,\mathrm{ns}\leq\Delta t_\mathrm{WP-CP}\leq125\,\mathrm{ns}$ with respect to the control pulse. For a given point with delay $\Delta t$ from the control pulse, there are at least $N=5$ wave packets with different $\Delta t_\mathrm{WP-CP}$, for which the point is contained deep inside the wave packet. For the $i-$th wave packet of the $N$ wave packets, we determine the instantaneous frequency $\omega_{\mathrm{inst,}i}(\Delta t)$ at the point $\Delta t$. Finally, we calculate $\omega_\mathrm{inst}(\Delta t)$ from $\omega_{\mathrm{inst,}i}(\Delta t)$ analogously to the case of the long wave packet discussed above and plot it as a white line in Fig.\,\ref{FIG:Fig3}.\\

\noindent\textbf{\large Data availability}\\
\noindent The measurement data presented in this article are available from the corresponding author upon reasonable request. Furthermore, these datasets supporting the findings of this study will be made available via Zenodo.\\

\bibliography{biblio}
\vspace{1cm}
\noindent\textbf{\large Acknowledgements}\\
\noindent F.A., N.C., and F.M.~acknowledge financial support from the European Union's Horizon Europe projects MiSS (Project ID: 101135868) and Qu-Pilot SGA (Project ID: 101113983).
The authors acknowledge support from Q@TN, the joint lab between the University of Trento, FBK, INFN, and CNR, financed by the Provincia Autonoma di Trento (PAT).
F.A., I.C., and F.M.~acknowledge support from the National Quantum Science and Technology Institute through the PNRR MUR Project under Grant PE0000023-NQSTI, co-funded by the European Union - NextGeneration EU.
G.R. and I.C.~acknowledge financial support from the Provincia Autonoma di Trento (PAT).
A.G.~acknowledges support from the Horizon 2020 Marie Sklodowska-Curie actions (H2020-MSCA-IF GA No.101027746).
A.G.~and F.M.~acknowledge support from the DARTWARS project funded by INFN within the Technological and Interdisciplinary Research Commission (CSN5).
I.C.~is grateful to Z.~Gaburro, M.~Ghulinyan, F.~Riboli, L.~Pavesi, and A.~Recati with whom the idea of photon energy lifting was originally conceived. A.V. and R.M.~thank G.~Richoux for her contribution to data acquisition. E.B.~thanks R.~Carobene for support with the RFSoC software.
The authors thank FBK cleanroom staff for support with the microfabrication of the device. \\

\noindent\textbf{\large Author contributions}\\
\noindent F.A., I.C., N.C., F.M., G.R., and A.V.~conceptualized the idea of dynamic Doppler frequency conversion in superconducting transmission lines. F.A., N.C., F.M., and A.V.~conceived the experimental implementation. The device was designed by A.G., while F.A., A.I., and F.M.~performed the microfabrication. F.A., E.B., R.M., and A.V.~performed the cryogenic measurements, while F.A., E.B., and R.M.~analyzed the data with input from I.C., N.C., F.M., G.R., and A.V. The manuscript was drafted by F.A., I.C., F.M., and G.R.~with the input of all authors.\\

\noindent\textbf{\large Competing interests}\\
\noindent F.A., E.B., N.C., A.I., and F.M.~are inventors on two patent applications (Italian application number 102025000010179 and international application number PCT/IB2025/058677) submitted by Fondazione Bruno Kessler related to dynamic-Doppler frequency-shifting devices utilizing high-kinetic-inductance transmission lines and Josephson-Junction-based transmission lines, respectively.

\setcounter{figure}{0}

\edfig{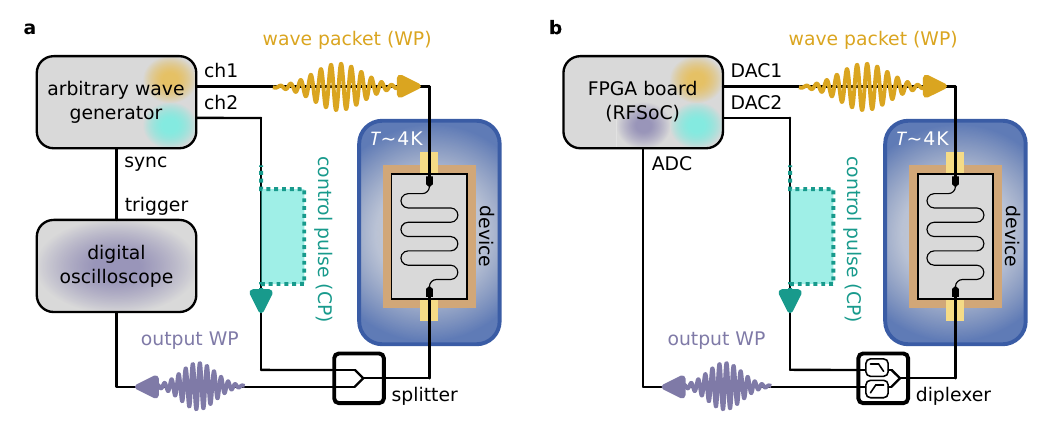}{0.8\linewidth}{\textbf{Experimental setups}. \textbf{a}, schematic layout of the experimental setup used to acquire full-trace measurements with a carrier frequency of 500\,MHz. \textbf{b}, schematic layout of the experimental setup used to acquire digitally down-converted data for wave packets with a carrier frequency of 4\,GHz.}{FIG:experimental_setup_extended}

\edfig{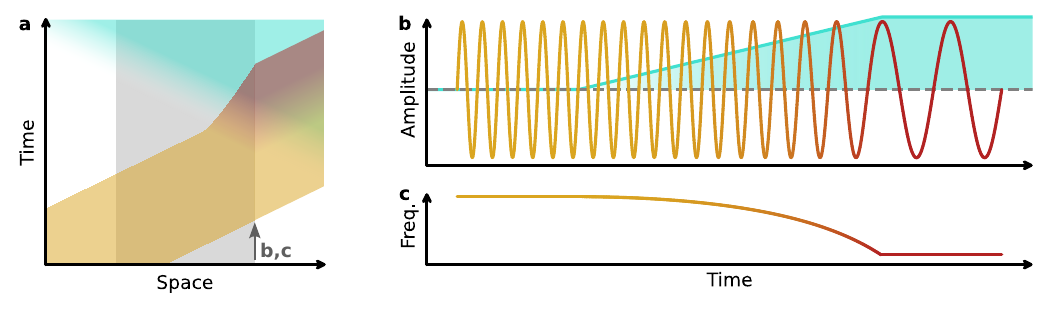}{0.8\linewidth}{\textbf{Instantaneous frequency shift.} \textbf{a}, Schematic spacetime diagram of a wave packet (orange-red) encountering a linearly increasing current front (white-turquoise). \textbf{b}, fixed-position cut of (\textbf{a}) at the output of the device showing the expected time evolution of both the wave packet (orange-red line) and the control-pulse waveform (turquoise line). \textbf{c}, time evolution of the wave packet's frequency.}{FIG:partial_lifting}

\edfig{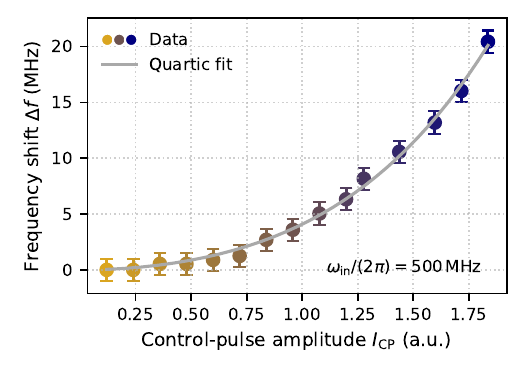}{0.49\linewidth}{\textbf{Amplitude-controlled frequency shift.} Measured blueshift of wave packets with an initial frequency $\omega_\mathrm{in}/(2\pi)=500\,\mathrm{MHz}$ after having traversed a falling control-pulse front of height $I_\mathrm{CP}$ inside the device. The utilized measurement setup is shown in Extended Data Fig.\,\ref{FIG:experimental_setup_extended}\textbf{a}. The grey line represents a polynomial fit to the data with quadratic and quartic contributions in $I_\mathrm{CP}$.}{FIG:shift_oscilloscope}

\end{document}